\documentclass[pra,twocolumn,showpacs,superscriptaddress,preprintnumbers,amsmath,amssymb,aps,floatfix]{revtex4-2}
\usepackage{mathrsfs}
\usepackage{amsmath}
\usepackage{amssymb}
\usepackage{graphicx}
\usepackage{dcolumn}
\usepackage{bm}
\usepackage{subfigure}
\usepackage{color}
\usepackage{ifpdf}
\usepackage{epstopdf}
\usepackage{verbatim}
\usepackage{bbm}
\usepackage{multirow}
\usepackage{booktabs}
\usepackage{color}
\usepackage[
pdfstartview=FitH,%
bookmarks=true,%
bookmarksopen=true,%
breaklinks=true,%
colorlinks=true,%
linkcolor=blue,anchorcolor=blue,%
citecolor=blue,filecolor=blue,%
menucolor=blue,%
urlcolor=blue]{hyperref}
\usepackage{float}
\usepackage[toc,page,title,titletoc,header]{appendix}
\usepackage{ragged2e}
\usepackage{lipsum}
\usepackage{physics}
\usepackage{array}
\usepackage{booktabs}

\usepackage{color}

\def\be{\begin{equation}}
	\def\ee{\end{equation}}
\def\bea{\begin{eqnarray}}
	\def\eea{\end{eqnarray}}

\begin{document}

\title{Influence of thermal effects on atomic Bloch oscillation}
\author{Guoling Yin}
\affiliation{State Key Laboratory of Quantum Optics and Quantum Optics Devices, Institute of Opto-Electronics, Shanxi University, Taiyuan 030006, China}
\author{Chi-Kin Lai}
\affiliation{State Key Laboratory of Advanced Optical Communication Systems and Networks, School of Electronics, Peking University, Beijing 100871, China}
\author{Nana Chang}
\affiliation{State Key Laboratory of Advanced Optical Communication Systems and Networks, School of Electronics, Peking University, Beijing 100871, China}
\affiliation{Institute of Carbon-based Thin Film Electronics, Peking University, Shanxi, Taiyuan 030012, China}
\author{Yi Liang}
\affiliation{State Key Laboratory of Advanced Optical Communication Systems and Networks, School of Electronics, Peking University, Beijing 100871, China}
\author{Dekai Mao}
\affiliation{State Key Laboratory of Advanced Optical Communication Systems and Networks, School of Electronics, Peking University, Beijing 100871, China}
\author{Xiaoji Zhou}\email{xjzhou@pku.edu.cn}
\affiliation{State Key Laboratory of Quantum Optics and Quantum Optics Devices, Institute of Opto-Electronics, Shanxi University, Taiyuan 030006, China}
\affiliation{State Key Laboratory of Advanced Optical Communication Systems and Networks, School of Electronics, Peking University, Beijing 100871, China}
\affiliation{Institute of Carbon-based Thin Film Electronics, Peking University, Shanxi, Taiyuan 030012, China}

\date{\today}
\begin{abstract}
Advancements in the experimental toolbox of cold atoms have enabled the meticulous control of atomic Bloch oscillation within optical lattices, thereby enhancing the capabilities of gravity interferometers. This work delves into the impact of thermal effects on Bloch oscillation in 1D accelerated optical lattices aligned with gravity by varying the system's initial temperature. Through the application of Raman cooling, we effectively reduce the longitudinal thermal effect, stabilizing the longitudinal coherence length over the timescale of its lifetime. The atomic losses over multiple Bloch oscillation is measured, which are primarily attributed to transverse excitation. Furthermore, we identify two distinct inverse scaling behaviors in the oscillation lifetime scaled by the corresponding density with respect to temperatures, implying diverse equilibrium processes within or outside the Bose-Einstein condensate regime. The competition between the system's coherence and atomic density leads to a relatively smooth variation in the actual lifetime versus temperature. Our findings provide valuable insights into the interaction between thermal effects and Bloch oscillation, offering avenues for the refinement of quantum measurement technologies.
\end{abstract}

\maketitle

\section{Introduction}
Atomic Bloch oscillation in optical lattices, manifesting as a dynamical oscillatory response to constant force~\cite{bloch1929quantenmechanik, zener1934theory}, has become a robust tool for quantum precision measurement. Compared with this phenomenon in solid-state materials, where rapid electron-electron and electron-impurity scattering sufficiently dampens the oscillation, optical lattices provide an impurity-free platform for exploring Bloch oscillation~\cite{PhysRevLett.76.4508, PhysRevA.55.2989, PhysRevLett.87.140402, Hartmann_2004,  PhysRevLett.100.080404, PhysRevLett.82.2022, 10.1063/1.881845}. The advancement in laser cooling techniques for cold atoms further facilitates the observation of the phenomenon by effectively loading atoms into optical lattices, either with direct loading of a Bose-Einstein condensate (BEC) or cold atoms combined with Raman cooling techniques~\cite{science.269.5221.198, PhysRevLett.75.3969, PhysRevLett.69.1741, PhysRevLett.75.4575, PhysRevA.70.043405}, while the former typically results in a denser atomic distribution where interaction plays the important role.

The oscillation behavior in optical lattices can be achieved by aligning the lattice along gravitational acceleration, providing a constant force that drives the oscillatory dynamics~\cite{Modugno2004AtomII, PhysRevLett.100.080404, PhysRevLett.92.230402, PhysRevLett.97.060402, Muller/science.aay6428}. This scheme can extend to the measurement of local gravitational acceleration, $g$, through the determination of the Bloch period $T_{\rm B}$ with ultracold atoms in optical lattices~\cite{PhysRevLett.97.060402, clade2005promising, PhysRevLett.106.038501}. Furthermore, these setups have been applied to direct measurements of the gravitational constant, $G$~\cite{rosi2014precision, tino2021, Kasevich2007}, the curvature of the gravitational field in ultracold atom systems~\cite{PhysRevLett.114.013001}, the fine structure constant~\cite{PhysRevA.74.052109, doi:10.1126/science.aap7706}, and examination of the equivalence principle~\cite{PhysRevLett.113.023005}. Moreover, the response of atoms to external force or acceleration also offers potential for high-sensitivity force measurements~\cite{GUO20222291}, Alternatively, the oscillation can be driven by inducing an inertial force on the atoms through the acceleration of the optical lattice~\cite{PhysRevLett.76.4508, PhysRevA.55.2989, PhysRevA.58.1480, Denschlag_2002}. In such acceleration-driven scenarios, atoms in different bands exhibit opposite group velocities, $v_g$~\cite{PhysRevA.107.023303}, and the adiabaticity may be constrained by a time bound that depends on the lattice's acceleration~\cite{PhysRevA.108.033310}.

Some studies have proposed utilizing Bloch oscillation in both acceleration-driven and gravity-driven setups to potentially enhance gravity interferometer capabilities~\cite{PhysRevA.88.031605, clade2015bloch, PhysRevA.85.013639, Bouchendira2012, Andia2015}, as the elongated path for atoms to fall freely leads to an extended interrogation time. However, the detailed thermal effects on these interferometers, which significantly affect their performance, have been less demonstrated. The exploration and understanding of the underlying thermal behaviors are critical, as they can substantially improve the sensitivity and accuracy of quantum precision measurements.

In this paper, we study Bloch oscillation in optical lattices driven by both gravity and acceleration, for atoms at different temperatures. Through systematic measurements of Bloch oscillation, we measure the coherence aligned with the lattice at different initial temperatures, indicating that the longitudinal coherence length extends for several lattice sites after applying the Raman cooling technique, and remains stable when increasing the evolution time in the lattice. The constant longitudinal coherence suggests the decay mechanism of Bloch oscillation is primarily linked to transverse excitation within the lattice. The measured lifetime of Bloch oscillation reveals an unexpected upward trend at different initial temperatures of the system, whereas the lifetime scaled by the average atomic density exhibits two distinct scaling against temperature, implying the varied thermal dynamics for atoms within or outside the BEC regime.

The structure of this paper is as follows. In Sec. \ref{sec:theoreticalModel}, we outline the theoretical model for Bloch oscillation under both gravity and acceleration-driven conditions. In Sec. \ref{sec:experimentImplementation}, we provide a comprehensive overview of the experimental setup and elucidate the process of Bloch oscillation within a 1D accelerated optical lattice. Experimental results, including observations of the stable longitudinal coherence length in Bloch oscillation and atomic lifetime of the oscillation relative to system temperature, are presented in Sec. \ref{sec:coherence length} and Sec. \ref{sec:lifetime}. We give a conclusion of our findings in Sec. \ref{sec:conclusion}.
\section{Theoretical model}\label{sec:theoreticalModel}

\begin{figure}[http]
    \includegraphics[width=0.45\textwidth]{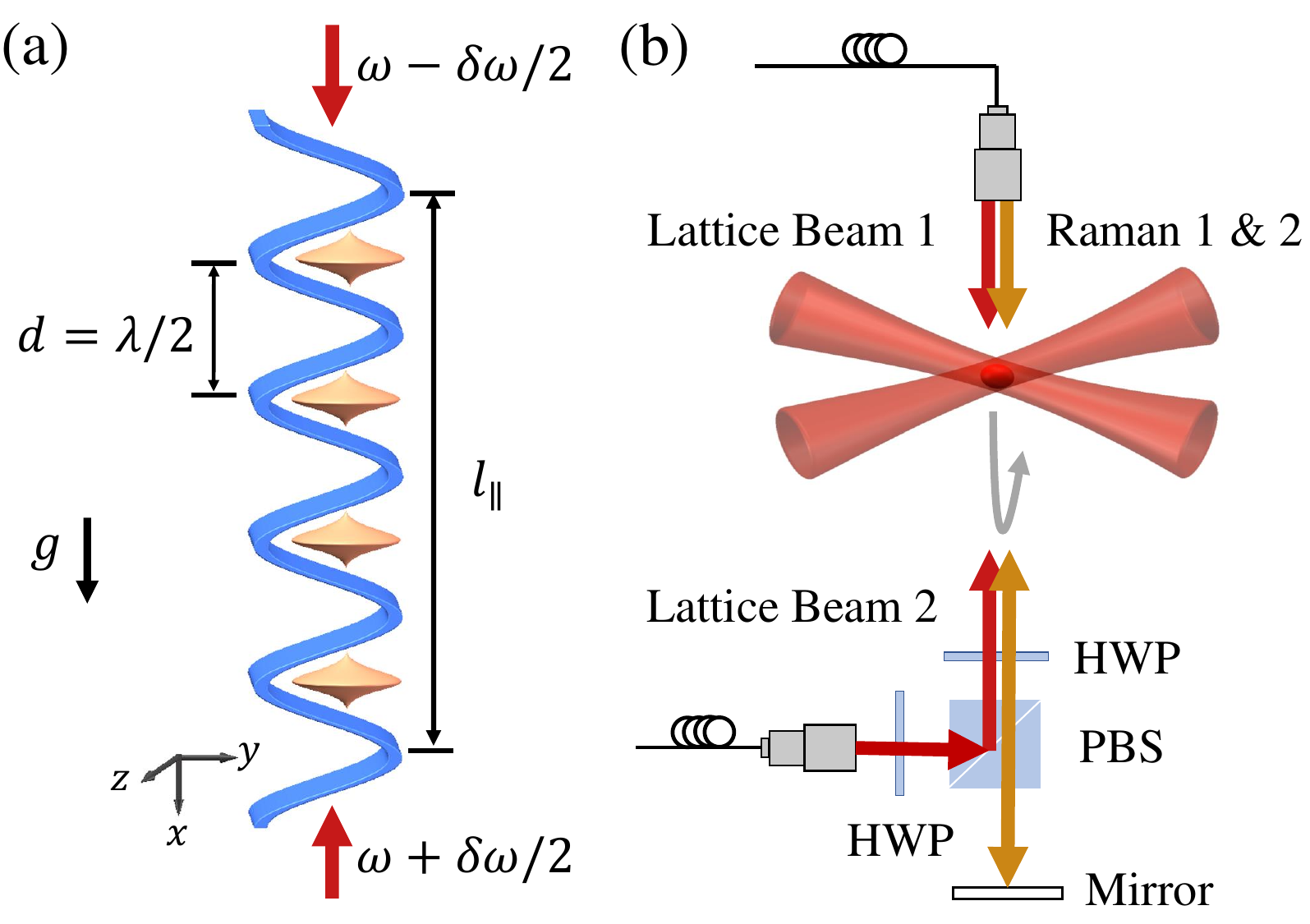}
    \caption{Experimental setup for both acceleration and gravity-driven Bloch oscillation. (a) Longitudinal movement of atoms in the 1D accelerated optical lattices under gravity, illustrating the periodic potential aligned with gravity in $x$-axis, depicted as a black arrow, and atoms with a pancake-shape spatial distribution. The acceleration of the optical lattice is achieved by modulating the frequency difference $\delta \omega=k\alpha t$ between the two lattice beams, as indicated by the two red arrows. In the co-moving frame, atoms are subject to a net force $m(g-\alpha)$. The longitudinal length $l_{\parallel}$ and lattice constant $d$ are displayed in the figure. (b) Schematic diagram of the setup, illustrating the atoms are prepared within a crossed 1064 nm optical dipole trap. It highlights the Raman and lattice beams in the vertical direction, including the counter-propagating configuration achieved by reflecting one of the Raman beams. The downward lattice beam is combined with Raman beams in the same optical fiber, while upward lattice beams are merged with one Raman beam using a polarizing beam splitter (PBS) and two half-wave plates (HWP).}
    \label{fig:figure1}
\label{fig:figureX}

\end{figure}

In an optical lattice system, we can modulate the frequency difference between two incident beams, resulting in the movement of the lattice. If the frequency difference is modulated by a constant acceleration $\alpha$, the lattice will be accelerated and move along the direction of the lattice. For our experiment, the lattice aligns with the direction of gravitational acceleration, $g$, leading to a gradient potential $-mg\hat{x}$ along the lattice, as shown in Fig.~\ref{fig:figure1}(a). Thus, the one dimensional Hamiltonian in the lab frame reads
\begin{align}
    \hat{H}^{\rm lab} = \frac{\hat{p}^2}{2m} + V_0\sin^{2}\left[k\left(x-\frac{1}{2}\alpha t^2\right)\right] - mg\hat{x},
\end{align}
where $V_0$ is the lattice depth, $k = \pi/d$ is the wave vector, $d$ is the lattice constant, $m$ is the atomic mass, and $\hat{p}$ is the momentum operator.

Considering the physical equivalence across different frames, analyzing in the co-moving frame is more convenient. In this frame, atoms experience an inertial force $-m\alpha$, which leads to the Hamiltonian:
\begin{align}
\hat{H}^{\rm cm} = \frac{\hat{p}^2}{2m} + V_0\sin^{2}(kx) - m(g-\alpha)\hat{x}.
\end{align}
To further analyse, we employ a unitary transformation:
\begin{align}
\hat{U}(t) = \exp\left(-\frac{\mathrm{i}}{\hbar}p_0(t)\hat{x}\right),
\end{align}
where $p_0(t) = m(g-\alpha)t$. After the transformation, the Hamiltonian becomes
\begin{align}
\Tilde{H}^{\rm cm} = \frac{[\hat{p}+\hbar q(t)]^2}{2m} + V_0\sin^{2}(kx),
\end{align}
where $q(t) = m(g-\alpha)t/\hbar$ is the quasimomentum in co-moving frame. This quasimomentum increases linearly under the constant acceleration $g-\alpha$, until it reaches the boundary of the first Brillouin zone, where Bragg scattering results in a change in quasimomentum by $2\hbar k$. In contrast, the momentum of atoms after $n$ times Bragg scattering occurrence is given by $p(t)=q(t)+2n\hbar k$.

The group velocity oscillation corresponding to the ground state, instantaneous S band, in the co-moving frame is given by:
\begin{align}
v^{\rm cm}_{\rm g}(t) = \frac{1}{\hbar}\frac{\partial E(q)}{\partial q} = \frac{2Jd}{\hbar}\sin\left(\frac{2\pi t}{T_{\rm B}}\right),
\end{align}
with $E(q)=-2J\cos(qd)$ representing the energy dispersion of the $S$ band under the tight-binding limit and $J$ being the nearest neighbor tunneling of the S band. The Bloch period $T_{\rm B}$ is given by:
\begin{align}
T_{\rm B} = \frac{2\pi\hbar}{m|g-\alpha|d}.
\end{align}
Furthermore, the group velocity of atoms in the lab frame can be obtained through Galilean transformation, i.e., $v^{\rm lab}_{\rm g}(t) = v^{\rm cm}_{\rm g}(t) + \alpha t$.
\section{Experimental demonstration of both gravity and acceleration driven Bloch oscillation}\label{sec:experimentImplementation}
\begin{figure}[http]
    \centering
    \includegraphics[width=0.45\textwidth]{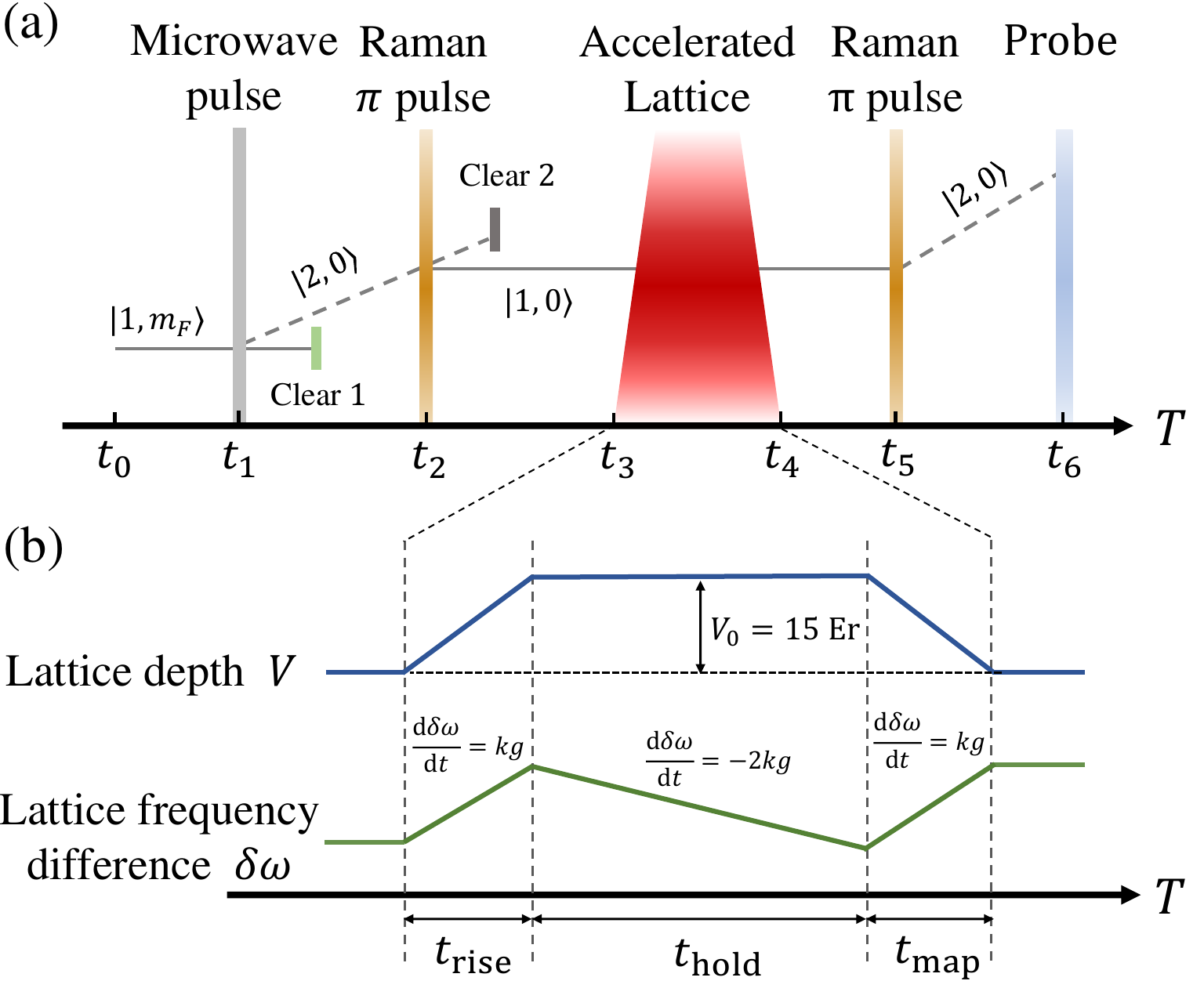}
    \caption{Experimental sequence for the accelerated optical lattice. (a) Initial state preparation begins at $t_0$ with atoms in $\ket{F = 1}$. At $t_1$, a microwave pulse transitions atoms to $\ket{F = 2, m_F = 0}$, while atoms still in $\ket{F = 1}$ are removed using a state-specific clear beam. The first Raman pulse at $t_2$ narrows the velocity distribution and transfers atoms to $\ket{F = 1, m_F = 0}$. Any remaining atoms in $\ket{F = 2}$ are then cleared. The optical lattice operation commences from $t_3$ to $t_4$, ending with a second Raman pulse at $t_5$ for velocity measurement by transferring resonant atoms to $\ket{F = 2, m_F = 0}$, followed by fluorescence detection at $t_6$. (b) The sequence for the accelerated optical lattice is detailed further: The lattice depth is linearly increased to $V_0=15$ Er within $t_{\rm rise}=1$ ms and is maintained during $t_{\rm hold}$. Phase modulation initiates by synchronizing the lattice's velocity with the atoms' free fall, adjusting the lattice frequency difference $\delta \omega$ at a rate of $kg$. To induce Bloch oscillation, $\delta \omega$ is modulated at a rate corresponding to $\alpha=-2g$. Finally, $\delta \omega$ is adjusted at a rate of $kg$ to ensure zero relative velocity between the lattice and the atoms, facilitating mapping of the atoms' states into quasimomentum space.}
    \label{fig:figure2}
\end{figure}

\begin{figure*}[http]
    \centering
    \includegraphics[width=1\textwidth]{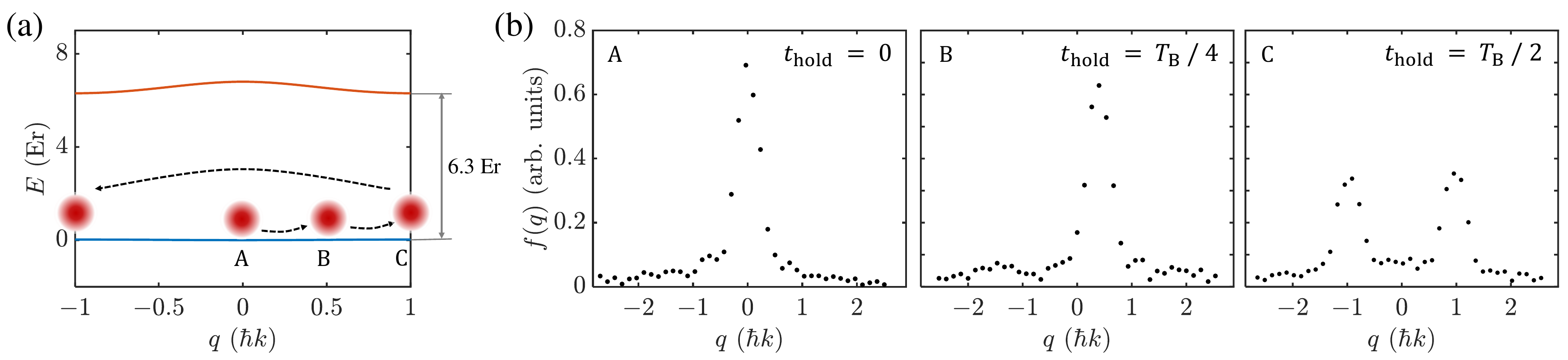}
    \caption{
    Bloch oscillation process in the co-moving frame. 
    (a) Illustration of Bloch oscillation (BO) driven by a constant acceleration $g-\alpha$ in the co-moving frame, showing atoms initially in the ground state ($q = 0$ at the instantaneous S band). Subject to a constant force $F=m(g-\alpha)$, the atoms' quasimomentum increases linearly until reaching the boundary of the first Brillouin zone. This induces Bragg scattering, resulting in quasimomentum shifts of $2\hbar k$. The BO process is depicted by black dashed arrows for the trajectory and red ellipses for atoms, corresponding to the momentum distribution of atoms at holding time: (A) $t_{\rm hold} = 0$, (B) $t_{\rm hold} = T_{\rm B}/4 = 99\ \mu s$, (C) $t_{\rm hold} = T_{\rm B}/2 = 198\ \mu s$.
    }
    \label{fig:figure3}
\end{figure*}
Our experiment is carried out in a 1D accelerated optical lattice aligned with the gravitational acceleration. Initially, we prepare approximately $1 \times 10^5$ $^{87}$Rb atoms in $\ket{F = 1}$ state, which is confined in the crossed 1064 nm optical dipole trap, as depicted in Fig.~\ref{fig:figure1}(b). The typical temperature of the system is 50 nK corresponds to harmonic trapping frequencies of $2\pi \times (29, 29, 41)$ Hz, which can be adjusted by altering the evaporation cooling parameters. Additionally, the system's initial temperature is determined through absorption imaging after a 20 ms time-of-flight (TOF). The critical temperature of the BEC phase transition, $T_c$, is around 200 nK. Upon switching off the optical dipole trap, the atoms fall freely. As illustrated in Fig.~\ref{fig:figure2}(a), the microwave pulse transfers atoms to $\ket{F = 2, m_F = 0}$, and the atoms remaining in $\ket{F = 1}$ state are blown by $\ket{F = 1}$ state clear beam. The Raman pulse, composed of two beams with a frequency difference of $6.8$ GHz, drives the Raman transition and selects a narrow velocity range of atoms. This process effectively limits the momentum broadening, $\delta q$, in the direction of the standing wave. The atomic state also transitions to $\ket{F = 1,m_F = 0}$ following the Raman pulse. Subsequently, the $\ket{F = 2}$ state clear beam blows the atom remaining in $\ket{F = 2}$. These processes not only transfer the atoms in a single $m_F$ state, but also ensure the momentum distribution is narrower than the first Brillouin zone, thereby effectively loading atoms into the optical lattice. The lattice itself is generated by two counter-propagating laser beams, with the frequency blue-detuned by $40 ~\mathrm{GHz}$ from the atomic transitions ($\lambda = 780.233~\mathrm{nm}$), minimizing the heating effect. Then, by modulating the frequency difference, $\delta \omega$, between the two lattice beams, the optical lattice is accelerated to load the atoms, achieve Bloch oscillation (BO), and map the atomic states into quasimomentum space. Finally, the atoms' quasimomentum distribution is analyzed by applying the second Raman pulse. Through sweeping the frequency of this pulse, the atoms with quasimomentum $q$ are resonant to the Raman frequency of $\omega=k_{\rm eff}q/m$, where $k_{\rm eff}=2k$ is the effective wave vector caused by the counter-propagating Raman beams. In the meanwhile, the resonant atoms with state $\ket{F = 1, m_F = 0}$ transition to $\ket{F = 2, m_F = 0}$. Therefore, the atoms' quasimomentum distribution, $f(q)$, can be delineated by the signal defined as $N_2/(N_1+N_2)$, where $N_i$ represents the atomic number in $\ket{F =i,m_F =0}$ state. The BO signal is measured through fluorescence detection following a 160 ms free fall until reaching the observation region.

The experimental sequence for the accelerated optical lattice, depicted in Fig.~\ref{fig:figure2}(b), is detailed as follows. Our method involves initially loading the atoms into the ground state of the moving periodic potential, namely, the instantaneous S band. To achieve this, the optical lattice is switched on by linearly increasing the lattice depth to 15 Er (Er denotes the recoil energy), which satisfies the tight-binding condition, over a rise time of $1~\mathrm{ms}$~\cite{PhysRevA.55.2989}. Concurrently, the frequency difference $\delta \omega$ between the two laser beams is modulated linearly via two acousto-optic modulators (AOM) at a constant rate of $\frac{\mathrm{d} \delta \omega}{\mathrm{d} t}=k\alpha=kg$, altering the lattice phase by $kgt^2/2$. This modulation enables the lattice to synchronize with the velocity of the freely falling atoms and adiabatically load them into the instantaneous S band. During this stage, BO does not occur due to the absence of relative acceleration between the atoms and the lattice. After the loading process, the lattice's acceleration is set to $\alpha = -2g$, allowing the atoms to ascend within the lattice, with their states evolving during the holding time $t_{\rm hold}$. In the co-moving frame, atoms are subjected to the combination of inertial force and gravity with $F = m(g-\alpha)=3mg$. Thus, BO is driven by a constant acceleration of $3g$. The BO process is observed by varying $t_{\rm hold}$ within the optical lattice. Subsequently, the lattice beams are linearly switched off over a $1~\mathrm{ms}$ duration, while simultaneously modulating $\delta \omega$ at a rate of $kg$ to maintain zero relative velocity between the lattice and the atoms. In this process, atoms populated in the instantaneous $S$ band are mapped to the first Brillouin zone. Therefore, the atoms' quasimomentum distribution can be measured after band mapping. 

Fig.~\ref{fig:figure3}(a) illustrates the Bloch oscillation process of atoms within the first Brillouin zone at a lattice depth of $V_0=15$ Er, where the energy gap between the $S$ and $P$ bands is $6.3$ Er, effectively suppressing Landau-Zener tunneling. Initially, atoms are loaded into the ground state. Subject to a constant external force in the co-moving frame, the atoms' quasimomentum linearly increases until it reaches the boundary of the first Brillouin zone. At this point, Bragg scattering propels the atoms to the opposite zone boundary, initiating a periodic motion as their quasimomentum continues to increase and cyclically returns to the starting point. Experimental observations of the Bloch oscillation process for various holding times $t_{\rm hold}$ are presented in Fig.~\ref{fig:figure3}(b). At $t_{\text{hold}} = 0$, atoms predominantly distribute around $q=0$. As $t_{\rm hold}$ extends, the atoms' quasimomentum linearly increases as a response of the constant force. At $t_{\rm hold}=98\ \rm\mu s$, the atoms' quasimomentum distribution peaks at $q=0.5\hbar k$, and the central peak of the distribution slightly decreases as atoms start to appear $2\hbar k$ away from $0.5\hbar k$. At $t_{\text{hold}} = 198\ \rm\mu s$, two symmetrical peaks positioning at $q = \pm \hbar k$ are observed, indicating Bragg scattering which occurs at $t_{\rm hold}=T_{\rm B}/2$. Thus, the Bloch period $T_{\rm B}$ is determined to be $396\ \rm\mu s$, which coincides with the oscillation period driven by a constant force of $3mg$, as anticipated by the theoretical model.
\section{Coherence analysis of Bloch oscillation along the optical lattice}\label{sec:coherence length}

\begin{figure}[http]
	\centering
	\includegraphics[width=0.4\textwidth]{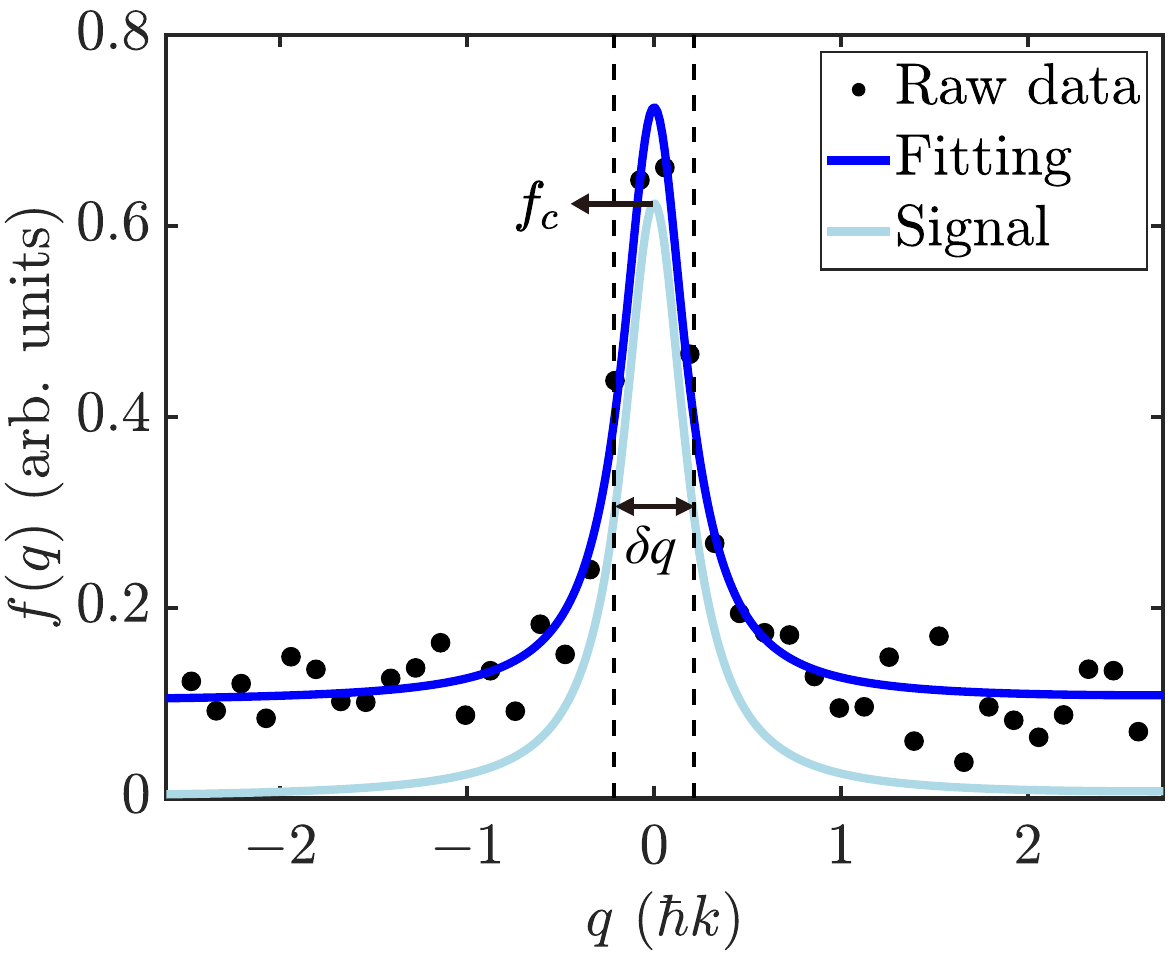}
	\caption{Fitting of the Bloch oscillation (BO) signal. Extraction of the quasimomentum distribution, $f(q)$, from the raw data (black dots) by fitting with a Lorentzian profile (blue solid line) at $t_{\rm hold}=30 T_{\rm B}$. The BO signal (light-blue solid line) is obtained by subtracting the fitted offset. The center of the signal, $q=0$, corresponds to a momentum amount of $2n\hbar k$ after $n$ occurrences of BO, and the central peak $f_c$ is denoted by a black arrow. The momentum broadening, $\delta q$, is extracted from the signal's full width at half maximum (FWHM).}
    \label{fig:figure4}
\end{figure}

\begin{figure}[http]
	\centering
    \includegraphics[width=0.42\textwidth]{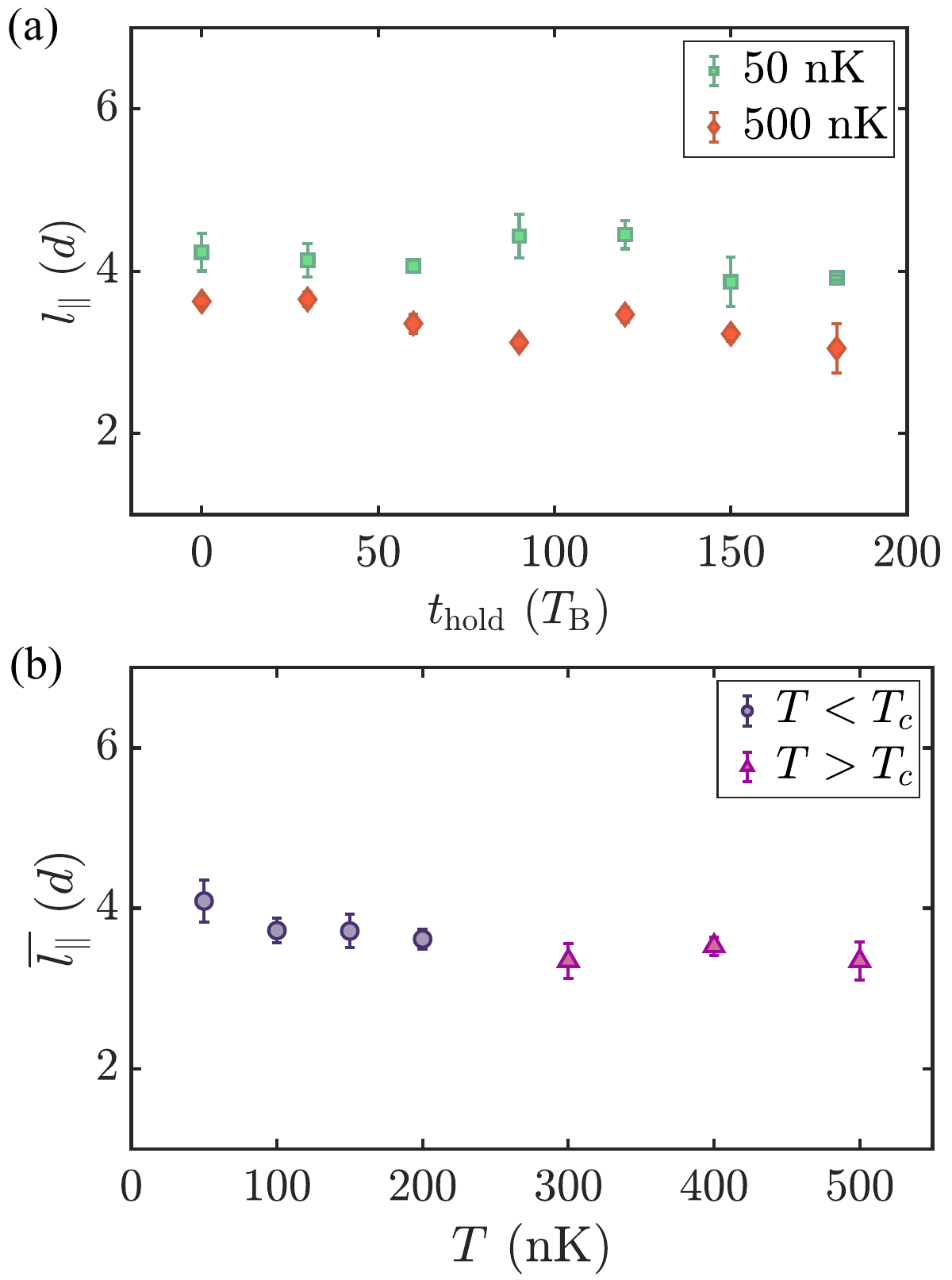}
	\caption{Measurement of the longitudinal coherence length. (a) The longitudinal coherence length, $l_{\parallel}$, for various initial temperatures, remains largely constant across increasing holding times, $t_{\rm hold}$, for atoms at 50 nK (green squares) and 500 nK (orange diamonds). The error bar represents the standard deviation of three repeated measurements. (b) The mean longitudinal coherence length, $\overline{l_{\parallel}}$, depicted as a function of initial temperature, shows data for atoms below (purple circles) and above (magenta triangles) the critical temperature $T_{c}$, averaged across various $t_{\rm hold}$ periods. Error bars indicate the corresponding standard deviations.}
    \label{fig:figure5}
\end{figure}

The longitudinal movement of atoms along the optical lattice ($x$-axis) can be effectively captured by the above theoretical model, particularly after applying Raman pulse aligned with the lattice. The pulse narrows the momentum broadening $\delta q$ along the lattice direction, serving as a cooling method to reduce the system's temperature. Consequently, the thermal effects in the longitudinal direction are significantly suppressed, leading to the enhancement of coherence of the system. 

To systematically analyse the coherence along the lattice, we measure the Bloch oscillation (BO) signal after multiple times of $T_{\rm B}$, and analyse the underlying coherence. Fig.~\ref{fig:figure4} displays the fitting process from typical BO signal for $t_{\rm hold}=30T_{\rm B}$. This signal, corresponding to the quasimomentum distribution $f(q)$, is analyzed by fitting it to a Lorentzian profile:
\begin{align}
    f(q) = \frac{A}{(q-q_0)^2 + (\delta q/2)^2} + B,
    \label{eq:Lorentzian}
\end{align}
where $\delta q$ denotes the momentum broadening, $q_0$ is the center of the quasimomentum peak, $A$ is the amplitude, and $B$ is the signal offset, corresponding to the incoherent fraction of atoms. The central peak of the quasimomentum distribution, $f_c = f(q_0) - B$, represents the atomic occupancy at the quasimomentum state $q_0$. When $t_{\rm hold} = nT_{\rm B}$, where $n$ is an integer, the distribution's peak occurs at $q_0 = 0$. This peak corresponds to a physical momentum of $p = 2n\hbar k$, indicating the atoms' return to their initial quasimomentum state after completing $n$ cycles of Bloch oscillation.

By adjusting the initial temperature of the system, we measure the longitudinal coherence length $l_{\parallel} = h / \delta q$ over various $t_{\rm hold}$ durations, as shown in Fig.~\ref{fig:figure5}(a). To ensure statistical reliability, each $l_{\parallel}$ value is measured three times, from which the mean and standard deviation are calculated. These observations indicate that $l_{\parallel}$ remains stable throughout the evolution in the optical lattice, up to a holding time of $t_{\rm hold} = 180T_{\rm B}$. This stability, observed in both the situation of atoms at $50$ nK and $500$ nK, underscores the effectiveness of Raman cooling in suppressing thermal effects along the lattice for different initial states, suggesting the coherence of atoms is not disrupted in the Bloch oscillation process. The mean longitudinal coherence length $\overline{l_{\parallel}}$, averaged over various holding times, is plotted as a function of the initial temperature in Fig.~\ref{fig:figure5}(b). A slight decrease in coherence length with rising temperature is observed, which can be attributed to the diluter atomic cloud before BEC formation, where Raman cooling becomes less effective above the critical temperature. However, the longitudinal coherence length reaches approximately four times the lattice constant, $d$, implying substantial long-range correlation aligned with the lattice.

\section{Atomic lifetime of Bloch oscillation at different initial temperature}\label{sec:lifetime}

In a one-dimensional optical lattice, especially within a deep lattice where $V_0 > 10\, \text{Er}$, atoms typically form a pancake-shaped distribution, with tunneling along the lattice direction significantly suppressed. Collisions of atoms mainly occur in the plane perpendicular to the lattice ($y-z$ plane), making transverse movement the principal factor in the decay mechanism. Consequently, transverse excitation leads to the damping of Bloch oscillation, with atoms potentially scattering out of the lattice's pancake-shaped distribution. The maintained coherence in the longitudinal direction further suggests that the reduction of the central peak $f_{\rm c}$ in the quasimomentum distribution over multiples of BO is primarily due to atomic losses in the transverse, unconfined direction, underscoring transverse excitation as the key decay mechanism. The direct observation of transverse momentum distribution, especially in a setup involving a free-falling and ascending process, is challenging. Nonetheless, the depletion in $f_{\rm c}$ over $t_{\rm hold}$ reveals the significance of the decay mechanism.

\begin{figure}[http]
	\centering
	\includegraphics[width=0.42\textwidth]{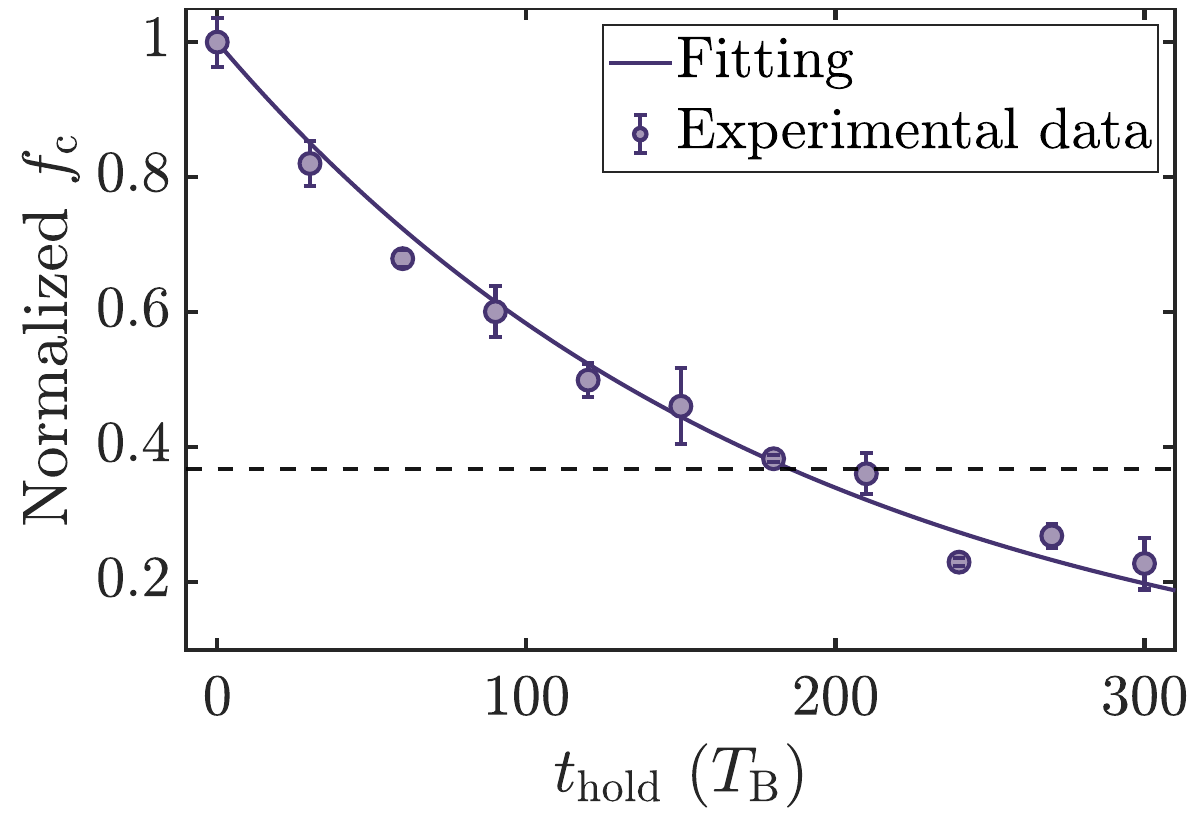}
    \caption{Evaluation of Bloch oscillation lifetime. Lifetime values are derived from the central peak, $f_c$, of the quasimomentum distribution for various holding times, $t_{\rm hold}$. Data points for atoms at 50 nK (purple circles) are normalized to the value at $t_{\rm hold} = 0$. Error bars denote the standard deviation from three independent measurements. The calculated lifetime, $\tau$, is indicated by the point where a black dashed line at $\exp(-1)$ intersects the solid purple fitting curve.}
    \label{fig:figure6}
\end{figure}

\begin{figure}[http]
	\centering
	\includegraphics[width=0.42\textwidth]{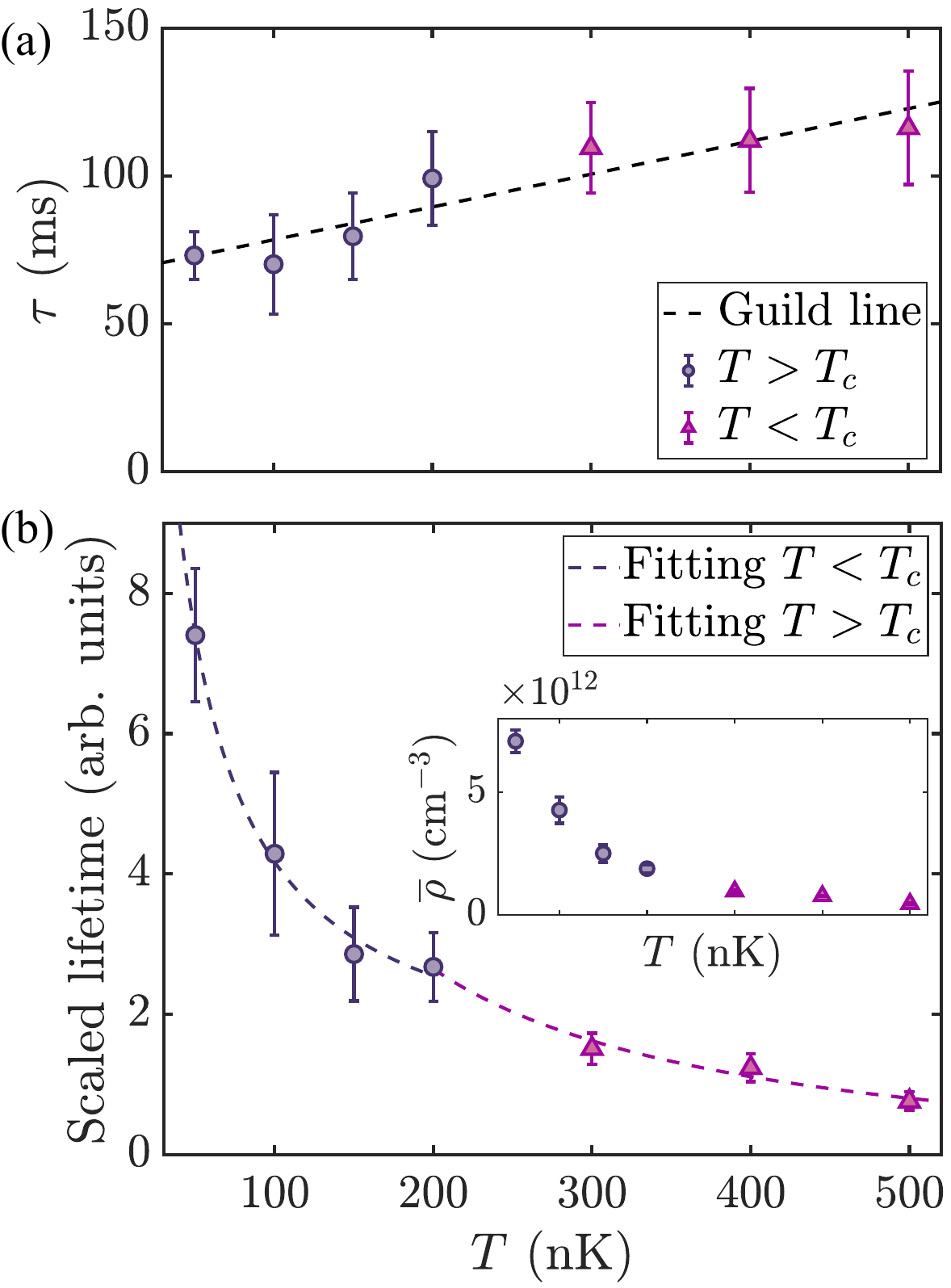}
    \caption{Lifetimes of BO at varying initial temperatures. (a) Lifetimes of BO across different temperatures for atoms below (purple circles) and above (magenta triangles) the critical temperature, $T_c$, with an upward trend highlighted by a black dashed guideline. Error bars correspond to fitting uncertainties. (b) Scaled lifetimes of Bloch oscillation, normalized by the initial average density $\overline{\rho}$ (illustrated in the inset), exhibit two distinct $T^{-1}$ scaling behaviors for conditions within and outside the BEC regime, depicted by purple ($T<T_c$) and magenta ($T>T_c$) dashed lines, respectively. Error bars are calculated by combining fitting uncertainties in $\tau$ with the standard deviation of $\overline{\rho}$ from three measurements.}
    \label{fig:figure7}
\end{figure}

To systematically investigate the decay process of BO, we set $t_{\text{hold}} = n T_{\text{B}}$ and examine how the lifetime varies with atoms' initial temperature. Fig.~\ref{fig:figure6} displays the lifetime fitting for the BO signal of atoms at 50 nK with the function 
\begin{align}
    f_{\rm c}(t) = f_{\rm c}(0)\exp(-t/\tau),
\end{align}
where $\tau$ is the BO lifetime. For system's initial temperature of 50 nK, the lifetime is calculated as \( \tau = 73.2 \pm 8.0 \) ms, corresponding to approximately 185 $T_{\rm B}$. Fig.~\ref{fig:figure7}(a) shows a counter-intuitive increase in BO lifetime versus system's initial temperature. The atomic lifetime of a driven one-dimensional optical lattice typically follows the relationship $\tau \propto \rho^{-1}$, where $\rho$ is the atomic density~\cite{PhysRevA.91.023624}. Thus, we measure the initial average atomic density $\overline{\rho}$ of the system across temperatures through absorption imaging. Notably, atoms at different temperatures generally show a large discrepancy in atomic density across the BEC phase transition; for instance, the density of atoms at 50 nK is roughly an order of magnitude higher than that of atoms at 500 nK, significantly influencing the atomic lifetime. To exclude the impact of atomic density, we scale the measured lifetime $\tau$ with the corresponding average atomic density $\overline{\rho}$ at different temperatures, as illustrated in Fig.~\ref{fig:figure7}(b). The scaled lifetime, indicative of the system's coherence after excluding scattering-induced decay, exhibits two distinct inverse scaling behaviors, represented by $T^{-1}$. This distinction implies dissimilar thermal dynamics among cold atoms with different temperatures subsequent to Raman cooling.

The counter-intuitive increase in lifetime with temperature, alongside divergent scaling behaviors, shows unexpected phenomena. These can be attributed to two primary factors. Firstly, the equilibrium state of atoms at $T<T_c$ is established through thermal exchange between the core BEC atoms and the surrounding non-condensed atoms, which are in different phases. In contrast, the equilibrium process for atoms at $T>T_c$ occurs within a single phase. This fundamental difference in how equilibrium is reached introduces disparate dynamics as the systems approach their final temperatures $T^{*}$ following Raman cooling, significantly influencing the system's coherence. As a result, the scaled lifetime with respect to initial temperature displays two different scaling. Secondly, while the direct measurement of BO lifetime exhibits a gradual increase with temperature, both the density and the scaled lifetime display a sharp transition at \(T = 200\) nK, where the BEC phase transition occurs. The interplay between atomic spatial distribution and the equilibrium process contributes to a relatively smooth shift in lifetime across the critical temperature. In other words, the abrupt transitions in both density and scaled lifetime linked to coherence counterbalance each other, resulting in a gradual increase in the physical lifetime with temperature. Nonetheless, these results leave room for more profound theoretical examination and detailed system modeling.

\section{Conclusion}\label{sec:conclusion}
In summary, we observe Bloch oscillation in 1D accelerated optical lattices along the gravity, spanning atoms with different temperatures. We demonstrate that Raman cooling effectively reduces longitudinal thermal effects, maintaining the stable longitudinal coherence length. Therefore, we attribute the observed atomic losses primarily to the transverse, unconstrained direction, caused by transverse excitation. Interestingly, while the lifetime of Bloch oscillation gradually increases with temperature, the lifetime scaled by density reveals two unique inverse scaling. This phenomenon demonstrates the significance of the equilibrium process in the decay mechanisms of Bloch oscillation and points to the intricate relationship between thermal dynamics and atomic density distribution. Our experimental results suggest that while a BEC offers a more suitable environment for exploring lattice dynamics and quantum simulations due to its enhanced coherence compared to non-condensed atoms. However, in certain situations, non-condensed atoms may offer advantages for quantum precision measurements due to their potentially longer lifetimes. These discoveries not only deepen our understanding of thermal effects in ultracold atoms systems but also pave the way for advancements in quantum measurement technologies.

\section*{Acknowledgements}
The authors would like to thank Z. Yu and L. Kong for useful discussions. This work is supported by the National Key Research and Development
Program of China (Grant Nos. 2021YFA0718300
and 2021YFA1400900), the National Natural Science Foundation of China (Grant Nos. 11920101004, 11934002, and 92365208), Science and Technology Major Project of Shanxi (Grant No. 202101030201022), and Space Application System of China Manned Space Program.
	

\bibliographystyle{apsrev}
\bibliography{AcceleratedOpticalLattices}

\begin{thebibliography}{38}
\expandafter\ifx\csname natexlab\endcsname\relax\def\natexlab#1{#1}\fi
\expandafter\ifx\csname bibnamefont\endcsname\relax
  \def\bibnamefont#1{#1}\fi
\expandafter\ifx\csname bibfnamefont\endcsname\relax
  \def\bibfnamefont#1{#1}\fi
\expandafter\ifx\csname citenamefont\endcsname\relax
  \def\citenamefont#1{#1}\fi
\expandafter\ifx\csname url\endcsname\relax
  \def\url#1{\texttt{#1}}\fi
\expandafter\ifx\csname urlprefix\endcsname\relax\def\urlprefix{URL }\fi
\providecommand{\bibinfo}[2]{#2}
\providecommand{\eprint}[2][]{\url{#2}}

\bibitem[{\citenamefont{Bloch}(1929)}]{bloch1929quantenmechanik}
\bibinfo{author}{\bibfnamefont{F.}~\bibnamefont{Bloch}},
  \bibinfo{journal}{Zeitschrift f{\"u}r physik} \textbf{\bibinfo{volume}{52}},
  \bibinfo{pages}{555} (\bibinfo{year}{1929}),
  \urlprefix\url{https://link.springer.com/article/10.1007/BF01339455}.

\bibitem[{\citenamefont{Zener}(1934)}]{zener1934theory}
\bibinfo{author}{\bibfnamefont{C.}~\bibnamefont{Zener}},
  \bibinfo{journal}{Proceedings of the Royal Society of London. Series A,
  Containing Papers of a Mathematical and Physical Character}
  \textbf{\bibinfo{volume}{145}}, \bibinfo{pages}{523} (\bibinfo{year}{1934}),
  \urlprefix\url{https://doi.org/10.1098/rspa.1934.0116}.

\bibitem[{\citenamefont{Ben~Dahan et~al.}(1996)\citenamefont{Ben~Dahan, Peik,
  Reichel, Castin, and Salomon}}]{PhysRevLett.76.4508}
\bibinfo{author}{\bibfnamefont{M.}~\bibnamefont{Ben~Dahan}},
  \bibinfo{author}{\bibfnamefont{E.}~\bibnamefont{Peik}},
  \bibinfo{author}{\bibfnamefont{J.}~\bibnamefont{Reichel}},
  \bibinfo{author}{\bibfnamefont{Y.}~\bibnamefont{Castin}}, \bibnamefont{and}
  \bibinfo{author}{\bibfnamefont{C.}~\bibnamefont{Salomon}},
  \bibinfo{journal}{Phys. Rev. Lett.} \textbf{\bibinfo{volume}{76}},
  \bibinfo{pages}{4508} (\bibinfo{year}{1996}),
  \urlprefix\url{https://link.aps.org/doi/10.1103/PhysRevLett.76.4508}.

\bibitem[{\citenamefont{Peik et~al.}(1997)\citenamefont{Peik, Ben~Dahan,
  Bouchoule, Castin, and Salomon}}]{PhysRevA.55.2989}
\bibinfo{author}{\bibfnamefont{E.}~\bibnamefont{Peik}},
  \bibinfo{author}{\bibfnamefont{M.}~\bibnamefont{Ben~Dahan}},
  \bibinfo{author}{\bibfnamefont{I.}~\bibnamefont{Bouchoule}},
  \bibinfo{author}{\bibfnamefont{Y.}~\bibnamefont{Castin}}, \bibnamefont{and}
  \bibinfo{author}{\bibfnamefont{C.}~\bibnamefont{Salomon}},
  \bibinfo{journal}{Phys. Rev. A} \textbf{\bibinfo{volume}{55}},
  \bibinfo{pages}{2989} (\bibinfo{year}{1997}),
  \urlprefix\url{https://link.aps.org/doi/10.1103/PhysRevA.55.2989}.

\bibitem[{\citenamefont{Morsch et~al.}(2001)\citenamefont{Morsch, M\"uller,
  Cristiani, Ciampini, and Arimondo}}]{PhysRevLett.87.140402}
\bibinfo{author}{\bibfnamefont{O.}~\bibnamefont{Morsch}},
  \bibinfo{author}{\bibfnamefont{J.~H.} \bibnamefont{M\"uller}},
  \bibinfo{author}{\bibfnamefont{M.}~\bibnamefont{Cristiani}},
  \bibinfo{author}{\bibfnamefont{D.}~\bibnamefont{Ciampini}}, \bibnamefont{and}
  \bibinfo{author}{\bibfnamefont{E.}~\bibnamefont{Arimondo}},
  \bibinfo{journal}{Phys. Rev. Lett.} \textbf{\bibinfo{volume}{87}},
  \bibinfo{pages}{140402} (\bibinfo{year}{2001}),
  \urlprefix\url{https://link.aps.org/doi/10.1103/PhysRevLett.87.140402}.

\bibitem[{\citenamefont{Hartmann et~al.}(2004)\citenamefont{Hartmann, Keck,
  Korsch, and Mossmann}}]{Hartmann_2004}
\bibinfo{author}{\bibfnamefont{T.}~\bibnamefont{Hartmann}},
  \bibinfo{author}{\bibfnamefont{F.}~\bibnamefont{Keck}},
  \bibinfo{author}{\bibfnamefont{H.~J.} \bibnamefont{Korsch}},
  \bibnamefont{and} \bibinfo{author}{\bibfnamefont{S.}~\bibnamefont{Mossmann}},
  \bibinfo{journal}{New Journal of Physics} \textbf{\bibinfo{volume}{6}},
  \bibinfo{pages}{2} (\bibinfo{year}{2004}),
  \urlprefix\url{https://dx.doi.org/10.1088/1367-2630/6/1/002}.

\bibitem[{\citenamefont{Gustavsson et~al.}(2008)\citenamefont{Gustavsson,
  Haller, Mark, Danzl, Rojas-Kopeinig, and N\"agerl}}]{PhysRevLett.100.080404}
\bibinfo{author}{\bibfnamefont{M.}~\bibnamefont{Gustavsson}},
  \bibinfo{author}{\bibfnamefont{E.}~\bibnamefont{Haller}},
  \bibinfo{author}{\bibfnamefont{M.~J.} \bibnamefont{Mark}},
  \bibinfo{author}{\bibfnamefont{J.~G.} \bibnamefont{Danzl}},
  \bibinfo{author}{\bibfnamefont{G.}~\bibnamefont{Rojas-Kopeinig}},
  \bibnamefont{and} \bibinfo{author}{\bibfnamefont{H.-C.}
  \bibnamefont{N\"agerl}}, \bibinfo{journal}{Phys. Rev. Lett.}
  \textbf{\bibinfo{volume}{100}}, \bibinfo{pages}{080404}
  (\bibinfo{year}{2008}),
  \urlprefix\url{https://link.aps.org/doi/10.1103/PhysRevLett.100.080404}.

\bibitem[{\citenamefont{Choi and Niu}(1999)}]{PhysRevLett.82.2022}
\bibinfo{author}{\bibfnamefont{D.-I.} \bibnamefont{Choi}} \bibnamefont{and}
  \bibinfo{author}{\bibfnamefont{Q.}~\bibnamefont{Niu}},
  \bibinfo{journal}{Phys. Rev. Lett.} \textbf{\bibinfo{volume}{82}},
  \bibinfo{pages}{2022} (\bibinfo{year}{1999}),
  \urlprefix\url{https://link.aps.org/doi/10.1103/PhysRevLett.82.2022}.

\bibitem[{\citenamefont{Raizen et~al.}(1997)\citenamefont{Raizen, Salomon, and
  Niu}}]{10.1063/1.881845}
\bibinfo{author}{\bibfnamefont{M.}~\bibnamefont{Raizen}},
  \bibinfo{author}{\bibfnamefont{C.}~\bibnamefont{Salomon}}, \bibnamefont{and}
  \bibinfo{author}{\bibfnamefont{Q.}~\bibnamefont{Niu}},
  \bibinfo{journal}{Physics Today} \textbf{\bibinfo{volume}{50}},
  \bibinfo{pages}{30} (\bibinfo{year}{1997}), ISSN \bibinfo{issn}{0031-9228},
  \urlprefix\url{https://doi.org/10.1063/1.881845}.

\bibitem[{\citenamefont{Anderson et~al.}(1995)\citenamefont{Anderson, Ensher,
  Matthews, Wieman, and Cornell}}]{science.269.5221.198}
\bibinfo{author}{\bibfnamefont{M.~H.} \bibnamefont{Anderson}},
  \bibinfo{author}{\bibfnamefont{J.~R.} \bibnamefont{Ensher}},
  \bibinfo{author}{\bibfnamefont{M.~R.} \bibnamefont{Matthews}},
  \bibinfo{author}{\bibfnamefont{C.~E.} \bibnamefont{Wieman}},
  \bibnamefont{and} \bibinfo{author}{\bibfnamefont{E.~A.}
  \bibnamefont{Cornell}}, \bibinfo{journal}{Science}
  \textbf{\bibinfo{volume}{269}}, \bibinfo{pages}{198} (\bibinfo{year}{1995}),
  \urlprefix\url{https://www.science.org/doi/abs/10.1126/science.269.5221.198}.

\bibitem[{\citenamefont{Davis et~al.}(1995)\citenamefont{Davis, Mewes, Andrews,
  van Druten, Durfee, Kurn, and Ketterle}}]{PhysRevLett.75.3969}
\bibinfo{author}{\bibfnamefont{K.~B.} \bibnamefont{Davis}},
  \bibinfo{author}{\bibfnamefont{M.~O.} \bibnamefont{Mewes}},
  \bibinfo{author}{\bibfnamefont{M.~R.} \bibnamefont{Andrews}},
  \bibinfo{author}{\bibfnamefont{N.~J.} \bibnamefont{van Druten}},
  \bibinfo{author}{\bibfnamefont{D.~S.} \bibnamefont{Durfee}},
  \bibinfo{author}{\bibfnamefont{D.~M.} \bibnamefont{Kurn}}, \bibnamefont{and}
  \bibinfo{author}{\bibfnamefont{W.}~\bibnamefont{Ketterle}},
  \bibinfo{journal}{Phys. Rev. Lett.} \textbf{\bibinfo{volume}{75}},
  \bibinfo{pages}{3969} (\bibinfo{year}{1995}),
  \urlprefix\url{https://link.aps.org/doi/10.1103/PhysRevLett.75.3969}.

\bibitem[{\citenamefont{Kasevich and Chu}(1992)}]{PhysRevLett.69.1741}
\bibinfo{author}{\bibfnamefont{M.}~\bibnamefont{Kasevich}} \bibnamefont{and}
  \bibinfo{author}{\bibfnamefont{S.}~\bibnamefont{Chu}},
  \bibinfo{journal}{Phys. Rev. Lett.} \textbf{\bibinfo{volume}{69}},
  \bibinfo{pages}{1741} (\bibinfo{year}{1992}),
  \urlprefix\url{https://link.aps.org/doi/10.1103/PhysRevLett.69.1741}.

\bibitem[{\citenamefont{Reichel et~al.}(1995)\citenamefont{Reichel, Bardou,
  Dahan, Peik, Rand, Salomon, and Cohen-Tannoudji}}]{PhysRevLett.75.4575}
\bibinfo{author}{\bibfnamefont{J.}~\bibnamefont{Reichel}},
  \bibinfo{author}{\bibfnamefont{F.}~\bibnamefont{Bardou}},
  \bibinfo{author}{\bibfnamefont{M.~B.} \bibnamefont{Dahan}},
  \bibinfo{author}{\bibfnamefont{E.}~\bibnamefont{Peik}},
  \bibinfo{author}{\bibfnamefont{S.}~\bibnamefont{Rand}},
  \bibinfo{author}{\bibfnamefont{C.}~\bibnamefont{Salomon}}, \bibnamefont{and}
  \bibinfo{author}{\bibfnamefont{C.}~\bibnamefont{Cohen-Tannoudji}},
  \bibinfo{journal}{Phys. Rev. Lett.} \textbf{\bibinfo{volume}{75}},
  \bibinfo{pages}{4575} (\bibinfo{year}{1995}),
  \urlprefix\url{https://link.aps.org/doi/10.1103/PhysRevLett.75.4575}.

\bibitem[{\citenamefont{Boyer et~al.}(2004)\citenamefont{Boyer, Lising,
  Rolston, and Phillips}}]{PhysRevA.70.043405}
\bibinfo{author}{\bibfnamefont{V.}~\bibnamefont{Boyer}},
  \bibinfo{author}{\bibfnamefont{L.~J.} \bibnamefont{Lising}},
  \bibinfo{author}{\bibfnamefont{S.~L.} \bibnamefont{Rolston}},
  \bibnamefont{and} \bibinfo{author}{\bibfnamefont{W.~D.}
  \bibnamefont{Phillips}}, \bibinfo{journal}{Phys. Rev. A}
  \textbf{\bibinfo{volume}{70}}, \bibinfo{pages}{043405}
  (\bibinfo{year}{2004}),
  \urlprefix\url{https://link.aps.org/doi/10.1103/PhysRevA.70.043405}.

\bibitem[{\citenamefont{Modugno et~al.}(2004)\citenamefont{Modugno,
  de~Mirand{\'e}s, Ferlaino, Ott, Roati, and Inguscio}}]{Modugno2004AtomII}
\bibinfo{author}{\bibfnamefont{G.}~\bibnamefont{Modugno}},
  \bibinfo{author}{\bibfnamefont{E.}~\bibnamefont{de~Mirand{\'e}s}},
  \bibinfo{author}{\bibfnamefont{F.}~\bibnamefont{Ferlaino}},
  \bibinfo{author}{\bibfnamefont{H.}~\bibnamefont{Ott}},
  \bibinfo{author}{\bibfnamefont{G.}~\bibnamefont{Roati}}, \bibnamefont{and}
  \bibinfo{author}{\bibfnamefont{M.}~\bibnamefont{Inguscio}},
  \bibinfo{journal}{Fortschritte der Physik} \textbf{\bibinfo{volume}{52}}
  (\bibinfo{year}{2004}),
  \urlprefix\url{https://api.semanticscholar.org/CorpusID:15240958}.

\bibitem[{\citenamefont{Roati et~al.}(2004)\citenamefont{Roati, de~Mirandes,
  Ferlaino, Ott, Modugno, and Inguscio}}]{PhysRevLett.92.230402}
\bibinfo{author}{\bibfnamefont{G.}~\bibnamefont{Roati}},
  \bibinfo{author}{\bibfnamefont{E.}~\bibnamefont{de~Mirandes}},
  \bibinfo{author}{\bibfnamefont{F.}~\bibnamefont{Ferlaino}},
  \bibinfo{author}{\bibfnamefont{H.}~\bibnamefont{Ott}},
  \bibinfo{author}{\bibfnamefont{G.}~\bibnamefont{Modugno}}, \bibnamefont{and}
  \bibinfo{author}{\bibfnamefont{M.}~\bibnamefont{Inguscio}},
  \bibinfo{journal}{Phys. Rev. Lett.} \textbf{\bibinfo{volume}{92}},
  \bibinfo{pages}{230402} (\bibinfo{year}{2004}),
  \urlprefix\url{https://link.aps.org/doi/10.1103/PhysRevLett.92.230402}.

\bibitem[{\citenamefont{Ferrari et~al.}(2006)\citenamefont{Ferrari, Poli,
  Sorrentino, and Tino}}]{PhysRevLett.97.060402}
\bibinfo{author}{\bibfnamefont{G.}~\bibnamefont{Ferrari}},
  \bibinfo{author}{\bibfnamefont{N.}~\bibnamefont{Poli}},
  \bibinfo{author}{\bibfnamefont{F.}~\bibnamefont{Sorrentino}},
  \bibnamefont{and} \bibinfo{author}{\bibfnamefont{G.~M.} \bibnamefont{Tino}},
  \bibinfo{journal}{Phys. Rev. Lett.} \textbf{\bibinfo{volume}{97}},
  \bibinfo{pages}{060402} (\bibinfo{year}{2006}),
  \urlprefix\url{https://link.aps.org/doi/10.1103/PhysRevLett.97.060402}.

\bibitem[{\citenamefont{Xu et~al.}(2019)\citenamefont{Xu, Jaffe, Panda,
  Kristensen, Clark, and Müller}}]{Muller/science.aay6428}
\bibinfo{author}{\bibfnamefont{V.}~\bibnamefont{Xu}},
  \bibinfo{author}{\bibfnamefont{M.}~\bibnamefont{Jaffe}},
  \bibinfo{author}{\bibfnamefont{C.~D.} \bibnamefont{Panda}},
  \bibinfo{author}{\bibfnamefont{S.~L.} \bibnamefont{Kristensen}},
  \bibinfo{author}{\bibfnamefont{L.~W.} \bibnamefont{Clark}}, \bibnamefont{and}
  \bibinfo{author}{\bibfnamefont{H.}~\bibnamefont{Müller}},
  \bibinfo{journal}{Science} \textbf{\bibinfo{volume}{366}},
  \bibinfo{pages}{745} (\bibinfo{year}{2019}),
  \urlprefix\url{https://www.science.org/doi/abs/10.1126/science.aay6428}.

\bibitem[{\citenamefont{Clad{\'e} et~al.}(2005)\citenamefont{Clad{\'e},
  Guellati-Kh{\'e}lifa, Schwob, Nez, Julien, and Biraben}}]{clade2005promising}
\bibinfo{author}{\bibfnamefont{P.}~\bibnamefont{Clad{\'e}}},
  \bibinfo{author}{\bibfnamefont{S.}~\bibnamefont{Guellati-Kh{\'e}lifa}},
  \bibinfo{author}{\bibfnamefont{C.}~\bibnamefont{Schwob}},
  \bibinfo{author}{\bibfnamefont{F.}~\bibnamefont{Nez}},
  \bibinfo{author}{\bibfnamefont{L.}~\bibnamefont{Julien}}, \bibnamefont{and}
  \bibinfo{author}{\bibfnamefont{F.}~\bibnamefont{Biraben}},
  \bibinfo{journal}{Europhysics Letters} \textbf{\bibinfo{volume}{71}},
  \bibinfo{pages}{730} (\bibinfo{year}{2005}),
  \urlprefix\url{https://iopscience.iop.org/article/10.1209/epl/i2005-10163-6}.

\bibitem[{\citenamefont{Poli et~al.}(2011)\citenamefont{Poli, Wang, Tarallo,
  Alberti, Prevedelli, and Tino}}]{PhysRevLett.106.038501}
\bibinfo{author}{\bibfnamefont{N.}~\bibnamefont{Poli}},
  \bibinfo{author}{\bibfnamefont{F.-Y.} \bibnamefont{Wang}},
  \bibinfo{author}{\bibfnamefont{M.~G.} \bibnamefont{Tarallo}},
  \bibinfo{author}{\bibfnamefont{A.}~\bibnamefont{Alberti}},
  \bibinfo{author}{\bibfnamefont{M.}~\bibnamefont{Prevedelli}},
  \bibnamefont{and} \bibinfo{author}{\bibfnamefont{G.~M.} \bibnamefont{Tino}},
  \bibinfo{journal}{Phys. Rev. Lett.} \textbf{\bibinfo{volume}{106}},
  \bibinfo{pages}{038501} (\bibinfo{year}{2011}),
  \urlprefix\url{https://link.aps.org/doi/10.1103/PhysRevLett.106.038501}.

\bibitem[{\citenamefont{Rosi et~al.}(2014)\citenamefont{Rosi, Sorrentino,
  Cacciapuoti, Prevedelli, and Tino}}]{rosi2014precision}
\bibinfo{author}{\bibfnamefont{G.}~\bibnamefont{Rosi}},
  \bibinfo{author}{\bibfnamefont{F.}~\bibnamefont{Sorrentino}},
  \bibinfo{author}{\bibfnamefont{L.}~\bibnamefont{Cacciapuoti}},
  \bibinfo{author}{\bibfnamefont{M.}~\bibnamefont{Prevedelli}},
  \bibnamefont{and} \bibinfo{author}{\bibfnamefont{G.}~\bibnamefont{Tino}},
  \bibinfo{journal}{Nature} \textbf{\bibinfo{volume}{510}},
  \bibinfo{pages}{518} (\bibinfo{year}{2014}),
  \urlprefix\url{https://www.nature.com/articles/nature13433}.

\bibitem[{\citenamefont{Tino}(2021)}]{tino2021}
\bibinfo{author}{\bibfnamefont{G.~M.} \bibnamefont{Tino}},
  \bibinfo{journal}{Quantum Science and Technology}
  \textbf{\bibinfo{volume}{6}}, \bibinfo{pages}{024014} (\bibinfo{year}{2021}),
  \urlprefix\url{https://dx.doi.org/10.1088/2058-9565/abd83e}.

\bibitem[{\citenamefont{Fixler et~al.}(2007)\citenamefont{Fixler, Foster,
  McGuirk, and Kasevich}}]{Kasevich2007}
\bibinfo{author}{\bibfnamefont{J.~B.} \bibnamefont{Fixler}},
  \bibinfo{author}{\bibfnamefont{G.~T.} \bibnamefont{Foster}},
  \bibinfo{author}{\bibfnamefont{J.~M.} \bibnamefont{McGuirk}},
  \bibnamefont{and} \bibinfo{author}{\bibfnamefont{M.~A.}
  \bibnamefont{Kasevich}}, \bibinfo{journal}{Science}
  \textbf{\bibinfo{volume}{315}}, \bibinfo{pages}{74} (\bibinfo{year}{2007}),
  \urlprefix\url{https://www.science.org/doi/abs/10.1126/science.1135459}.

\bibitem[{\citenamefont{Rosi et~al.}(2015)\citenamefont{Rosi, Cacciapuoti,
  Sorrentino, Menchetti, Prevedelli, and Tino}}]{PhysRevLett.114.013001}
\bibinfo{author}{\bibfnamefont{G.}~\bibnamefont{Rosi}},
  \bibinfo{author}{\bibfnamefont{L.}~\bibnamefont{Cacciapuoti}},
  \bibinfo{author}{\bibfnamefont{F.}~\bibnamefont{Sorrentino}},
  \bibinfo{author}{\bibfnamefont{M.}~\bibnamefont{Menchetti}},
  \bibinfo{author}{\bibfnamefont{M.}~\bibnamefont{Prevedelli}},
  \bibnamefont{and} \bibinfo{author}{\bibfnamefont{G.~M.} \bibnamefont{Tino}},
  \bibinfo{journal}{Phys. Rev. Lett.} \textbf{\bibinfo{volume}{114}},
  \bibinfo{pages}{013001} (\bibinfo{year}{2015}),
  \urlprefix\url{https://link.aps.org/doi/10.1103/PhysRevLett.114.013001}.

\bibitem[{\citenamefont{Clad\'e et~al.}(2006)\citenamefont{Clad\'e,
  de~Mirandes, Cadoret, Guellati-Kh\'elifa, Schwob, Nez, Julien, and
  Biraben}}]{PhysRevA.74.052109}
\bibinfo{author}{\bibfnamefont{P.}~\bibnamefont{Clad\'e}},
  \bibinfo{author}{\bibfnamefont{E.}~\bibnamefont{de~Mirandes}},
  \bibinfo{author}{\bibfnamefont{M.}~\bibnamefont{Cadoret}},
  \bibinfo{author}{\bibfnamefont{S.}~\bibnamefont{Guellati-Kh\'elifa}},
  \bibinfo{author}{\bibfnamefont{C.}~\bibnamefont{Schwob}},
  \bibinfo{author}{\bibfnamefont{F.~m.~c.} \bibnamefont{Nez}},
  \bibinfo{author}{\bibfnamefont{L.}~\bibnamefont{Julien}}, \bibnamefont{and}
  \bibinfo{author}{\bibfnamefont{F.~m.~c.} \bibnamefont{Biraben}},
  \bibinfo{journal}{Phys. Rev. A} \textbf{\bibinfo{volume}{74}},
  \bibinfo{pages}{052109} (\bibinfo{year}{2006}),
  \urlprefix\url{https://link.aps.org/doi/10.1103/PhysRevA.74.052109}.

\bibitem[{\citenamefont{Parker et~al.}(2018)\citenamefont{Parker, Yu, Zhong,
  Estey, and Müller}}]{doi:10.1126/science.aap7706}
\bibinfo{author}{\bibfnamefont{R.~H.} \bibnamefont{Parker}},
  \bibinfo{author}{\bibfnamefont{C.}~\bibnamefont{Yu}},
  \bibinfo{author}{\bibfnamefont{W.}~\bibnamefont{Zhong}},
  \bibinfo{author}{\bibfnamefont{B.}~\bibnamefont{Estey}}, \bibnamefont{and}
  \bibinfo{author}{\bibfnamefont{H.}~\bibnamefont{Müller}},
  \bibinfo{journal}{Science} \textbf{\bibinfo{volume}{360}},
  \bibinfo{pages}{191} (\bibinfo{year}{2018}),
  \eprint{https://www.science.org/doi/pdf/10.1126/science.aap7706},
  \urlprefix\url{https://www.science.org/doi/abs/10.1126/science.aap7706}.

\bibitem[{\citenamefont{Tarallo et~al.}(2014)\citenamefont{Tarallo, Mazzoni,
  Poli, Sutyrin, Zhang, and Tino}}]{PhysRevLett.113.023005}
\bibinfo{author}{\bibfnamefont{M.~G.} \bibnamefont{Tarallo}},
  \bibinfo{author}{\bibfnamefont{T.}~\bibnamefont{Mazzoni}},
  \bibinfo{author}{\bibfnamefont{N.}~\bibnamefont{Poli}},
  \bibinfo{author}{\bibfnamefont{D.~V.} \bibnamefont{Sutyrin}},
  \bibinfo{author}{\bibfnamefont{X.}~\bibnamefont{Zhang}}, \bibnamefont{and}
  \bibinfo{author}{\bibfnamefont{G.~M.} \bibnamefont{Tino}},
  \bibinfo{journal}{Phys. Rev. Lett.} \textbf{\bibinfo{volume}{113}},
  \bibinfo{pages}{023005} (\bibinfo{year}{2014}),
  \urlprefix\url{https://link.aps.org/doi/10.1103/PhysRevLett.113.023005}.

\bibitem[{\citenamefont{Guo et~al.}(2022)\citenamefont{Guo, Yu, Wei, Jin, Chen,
  Li, Zhang, and Zhou}}]{GUO20222291}
\bibinfo{author}{\bibfnamefont{X.}~\bibnamefont{Guo}},
  \bibinfo{author}{\bibfnamefont{Z.}~\bibnamefont{Yu}},
  \bibinfo{author}{\bibfnamefont{F.}~\bibnamefont{Wei}},
  \bibinfo{author}{\bibfnamefont{S.}~\bibnamefont{Jin}},
  \bibinfo{author}{\bibfnamefont{X.}~\bibnamefont{Chen}},
  \bibinfo{author}{\bibfnamefont{X.}~\bibnamefont{Li}},
  \bibinfo{author}{\bibfnamefont{X.}~\bibnamefont{Zhang}}, \bibnamefont{and}
  \bibinfo{author}{\bibfnamefont{X.}~\bibnamefont{Zhou}},
  \bibinfo{journal}{Science Bulletin} \textbf{\bibinfo{volume}{67}},
  \bibinfo{pages}{2291} (\bibinfo{year}{2022}), ISSN \bibinfo{issn}{2095-9273},
  \urlprefix\url{https://www.sciencedirect.com/science/article/pii/S2095927322004881}.

\bibitem[{\citenamefont{Berg-S\o{}rensen and
  M\o{}lmer}(1998)}]{PhysRevA.58.1480}
\bibinfo{author}{\bibfnamefont{K.}~\bibnamefont{Berg-S\o{}rensen}}
  \bibnamefont{and}
  \bibinfo{author}{\bibfnamefont{K.}~\bibnamefont{M\o{}lmer}},
  \bibinfo{journal}{Phys. Rev. A} \textbf{\bibinfo{volume}{58}},
  \bibinfo{pages}{1480} (\bibinfo{year}{1998}),
  \urlprefix\url{https://link.aps.org/doi/10.1103/PhysRevA.58.1480}.

\bibitem[{\citenamefont{Denschlag et~al.}(2002)\citenamefont{Denschlag,
  Simsarian, Häffner, McKenzie, Browaeys, Cho, Helmerson, Rolston, and
  Phillips}}]{Denschlag_2002}
\bibinfo{author}{\bibfnamefont{J.~H.} \bibnamefont{Denschlag}},
  \bibinfo{author}{\bibfnamefont{J.~E.} \bibnamefont{Simsarian}},
  \bibinfo{author}{\bibfnamefont{H.}~\bibnamefont{Häffner}},
  \bibinfo{author}{\bibfnamefont{C.}~\bibnamefont{McKenzie}},
  \bibinfo{author}{\bibfnamefont{A.}~\bibnamefont{Browaeys}},
  \bibinfo{author}{\bibfnamefont{D.}~\bibnamefont{Cho}},
  \bibinfo{author}{\bibfnamefont{K.}~\bibnamefont{Helmerson}},
  \bibinfo{author}{\bibfnamefont{S.~L.} \bibnamefont{Rolston}},
  \bibnamefont{and} \bibinfo{author}{\bibfnamefont{W.~D.}
  \bibnamefont{Phillips}}, \bibinfo{journal}{Journal of Physics B: Atomic,
  Molecular and Optical Physics} \textbf{\bibinfo{volume}{35}},
  \bibinfo{pages}{3095} (\bibinfo{year}{2002}),
  \urlprefix\url{https://dx.doi.org/10.1088/0953-4075/35/14/307}.

\bibitem[{\citenamefont{Yu et~al.}(2023)\citenamefont{Yu, Tian, Peng, Mao,
  Chen, and Zhou}}]{PhysRevA.107.023303}
\bibinfo{author}{\bibfnamefont{Z.}~\bibnamefont{Yu}},
  \bibinfo{author}{\bibfnamefont{J.}~\bibnamefont{Tian}},
  \bibinfo{author}{\bibfnamefont{P.}~\bibnamefont{Peng}},
  \bibinfo{author}{\bibfnamefont{D.}~\bibnamefont{Mao}},
  \bibinfo{author}{\bibfnamefont{X.}~\bibnamefont{Chen}}, \bibnamefont{and}
  \bibinfo{author}{\bibfnamefont{X.}~\bibnamefont{Zhou}},
  \bibinfo{journal}{Phys. Rev. A} \textbf{\bibinfo{volume}{107}},
  \bibinfo{pages}{023303} (\bibinfo{year}{2023}),
  \urlprefix\url{https://link.aps.org/doi/10.1103/PhysRevA.107.023303}.

\bibitem[{\citenamefont{Yin et~al.}(2023)\citenamefont{Yin, Kong, Yu, Tian,
  Chen, and Zhou}}]{PhysRevA.108.033310}
\bibinfo{author}{\bibfnamefont{G.}~\bibnamefont{Yin}},
  \bibinfo{author}{\bibfnamefont{L.}~\bibnamefont{Kong}},
  \bibinfo{author}{\bibfnamefont{Z.}~\bibnamefont{Yu}},
  \bibinfo{author}{\bibfnamefont{J.}~\bibnamefont{Tian}},
  \bibinfo{author}{\bibfnamefont{X.}~\bibnamefont{Chen}}, \bibnamefont{and}
  \bibinfo{author}{\bibfnamefont{X.}~\bibnamefont{Zhou}},
  \bibinfo{journal}{Phys. Rev. A} \textbf{\bibinfo{volume}{108}},
  \bibinfo{pages}{033310} (\bibinfo{year}{2023}),
  \urlprefix\url{https://link.aps.org/doi/10.1103/PhysRevA.108.033310}.

\bibitem[{\citenamefont{Andia et~al.}(2013)\citenamefont{Andia, Jannin, Nez,
  Biraben, Guellati-Kh\'elifa, and Clad\'e}}]{PhysRevA.88.031605}
\bibinfo{author}{\bibfnamefont{M.}~\bibnamefont{Andia}},
  \bibinfo{author}{\bibfnamefont{R.}~\bibnamefont{Jannin}},
  \bibinfo{author}{\bibfnamefont{F.~m.~c.} \bibnamefont{Nez}},
  \bibinfo{author}{\bibfnamefont{F.~m.~c.} \bibnamefont{Biraben}},
  \bibinfo{author}{\bibfnamefont{S.}~\bibnamefont{Guellati-Kh\'elifa}},
  \bibnamefont{and} \bibinfo{author}{\bibfnamefont{P.}~\bibnamefont{Clad\'e}},
  \bibinfo{journal}{Phys. Rev. A} \textbf{\bibinfo{volume}{88}},
  \bibinfo{pages}{031605} (\bibinfo{year}{2013}),
  \urlprefix\url{https://link.aps.org/doi/10.1103/PhysRevA.88.031605}.

\bibitem[{\citenamefont{Clad{\'e}}(2015)}]{clade2015bloch}
\bibinfo{author}{\bibfnamefont{P.}~\bibnamefont{Clad{\'e}}},
  \bibinfo{journal}{La Rivista del Nuovo Cimento}
  \textbf{\bibinfo{volume}{38}}, \bibinfo{pages}{173} (\bibinfo{year}{2015}),
  \urlprefix\url{https://doi.org/10.1393/ncr/i2015-10111-3}.

\bibitem[{\citenamefont{Charri\`ere et~al.}(2012)\citenamefont{Charri\`ere,
  Cadoret, Zahzam, Bidel, and Bresson}}]{PhysRevA.85.013639}
\bibinfo{author}{\bibfnamefont{R.}~\bibnamefont{Charri\`ere}},
  \bibinfo{author}{\bibfnamefont{M.}~\bibnamefont{Cadoret}},
  \bibinfo{author}{\bibfnamefont{N.}~\bibnamefont{Zahzam}},
  \bibinfo{author}{\bibfnamefont{Y.}~\bibnamefont{Bidel}}, \bibnamefont{and}
  \bibinfo{author}{\bibfnamefont{A.}~\bibnamefont{Bresson}},
  \bibinfo{journal}{Phys. Rev. A} \textbf{\bibinfo{volume}{85}},
  \bibinfo{pages}{013639} (\bibinfo{year}{2012}),
  \urlprefix\url{https://link.aps.org/doi/10.1103/PhysRevA.85.013639}.

\bibitem[{\citenamefont{Bouchendira}(2012)}]{Bouchendira2012}
\bibinfo{author}{\bibfnamefont{R.}~\bibnamefont{Bouchendira}},
  \bibinfo{type}{Thèse de doctorat}, \bibinfo{school}{Université Pierre et
  Marie Curie}, \bibinfo{address}{Paris} (\bibinfo{year}{2012}),
  \bibinfo{note}{soutenue publiquement le 17 Juillet 2012}.

\bibitem[{\citenamefont{Andia}(2015)}]{Andia2015}
\bibinfo{author}{\bibfnamefont{M.}~\bibnamefont{Andia}}, \bibinfo{type}{Thèse
  de doctorat}, \bibinfo{school}{Université Pierre et Marie Curie},
  \bibinfo{address}{Paris} (\bibinfo{year}{2015}), \bibinfo{note}{soutenue le
  25 septembre 2015}.

\bibitem[{\citenamefont{Choudhury and Mueller}(2015)}]{PhysRevA.91.023624}
\bibinfo{author}{\bibfnamefont{S.}~\bibnamefont{Choudhury}} \bibnamefont{and}
  \bibinfo{author}{\bibfnamefont{E.~J.} \bibnamefont{Mueller}},
  \bibinfo{journal}{Phys. Rev. A} \textbf{\bibinfo{volume}{91}},
  \bibinfo{pages}{023624} (\bibinfo{year}{2015}),
  \urlprefix\url{https://link.aps.org/doi/10.1103/PhysRevA.91.023624}.

\end{thebibliography}
 
\end{document}